\documentclass[onecolumn]{aastex62}
\usepackage{xspace}
\usepackage{xcolor}
\usepackage{graphicx}
\usepackage{rotating}
\usepackage{url}

\shorttitle{Southern Symbiotic Stars}
\shortauthors{Dickey, John M, et al.}

\begin{document}
\title{Radio Spectral Index Analysis of Southern Hemisphere Symbiotic Stars}
\author[0000-0002-6300-7459]{John M. Dickey}%\altaffiliation{1}
\affiliation{School of Natural Sciences, Private Bag 37, University of Tasmania, Hobart, TAS, 7001, Australia}
\author[0000-0001-7548-5266]{J. H. S. Weston}
\affiliation{Federated IT, 1201 Wilson Blvd, 27th Floor, Arlington VA 22209}
\author{J. L. Sokoloski} 
\affiliation{Astronomy Dept., Columbia University, New York, USA}
\affiliation{LSST Corporation, 933 N. Cherry Ave, Tucson, AZ 85721}
\author[0000-0002-7521-9897]{S.D. Vrtilek}
\affiliation{Harvard-Smithsonian Center for Astrophysics, 60 Garden Street, Cambridge, MA 02138, USA}
\author{Michael McCollough}
\affiliation{Harvard-Smithsonian Center for Astrophysics, 60 Garden Street, Cambridge, MA 02138, USA}

\received{2 December 2020}
\revised{27 January 2021}

\begin{abstract}
Symbiotic stars show emission across the electromagnetic spectrum from a wide array of physical processes. At cm-waves both synchrotron and thermal emission is seen, often highly variable and associated with outbursts in the optical and X-rays.  Most models of the radio emission include an ionized region within the dense wind of the red giant star, that is kept ionized by activity on the white dwarf companion or its accretion disk.  In some cases there is on-going shell burning on the white dwarf due to its high mass accretion rate or a prior nova eruption, in other cases nuclear fusion occurs only occasionally as recurrent nova events.  In this study we measure the spectral indices of a sample of symbiotic systems in the Southern Hemisphere using the Australia Telescope Compact Array.  Putting our data together with results from other surveys, we derive the optical depths and brightness temperatures of some well-known symbiotic stars.  Using parallax distances from Gaia Data Release 3, we determine the sizes and characteristic electron densities in the radio emission regions.  The results show a range of a factor of 10$^4$ in radio luminosity, and a factor of 100 in linear size.  These numbers are consistent with a picture where the rate of shell burning on the white dwarf determines the radio luminosity.  Therefore, our findings also suggest that radio luminosity can be used to determine whether a symbiotic star is powered by accretion alone or also by shell burning.

\end{abstract}


\section{Introduction}\label{sec:Introduction}
Mass-exchange binary stars demonstrate a tremendous diversity of emission processes.  When one of
the stars is a neutron star or black hole, the accretion process commonly produces hard X-rays as well as
synchrotron emitting jets, similar to those seen in quasars, hence the term "micro-quasar" \citep{Mirabel_1993}.  When the system contains a white dwarf with a red-giant companion, there is still a dramatic range of 
behavior observed, with the system often described as a symbiotic star (SySt) \citep{Merrill_1941}, see \citet{Munari_2019}
for a recent review. The classification was originally based on the presence of optical spectral
lines indicating a cool giant star mixed with very high ionization lines known at the time only in O stars
and planetary nebulae.  Since then symbiotic stars have
demonstrated emission from the radio through to $\gamma$ rays, often showing strong variability.  Many of these
properties are now identified with the accretion onto the white dwarf and the resulting shell burning, or
nuclear fusion on the surface of the star.  In some cases neutron star binaries are included in catalogs of
SySts \citep[type $\gamma$]{Belczynski_etal_2000,vandenEijnden_etal_2018,Luna_etal_2013}.  Conditions in the red giant wind, and in particular its outflow velocity, can be determined from the optical spectrum, that often shows 
absorption lines or molecular bands such as TiO, H$_2$O, CO, CN and VO.
These are in striking contrast to the 
high ionization species, e.g. [O {\tt{VI}}] and other indicators of high temperature gas such as X-ray or $\gamma$ ray emission \citep{Giroletti_etal_2020}.  The X-ray spectrum can often distinguish between systems 
where the hot gas is in the boundary between the accretion disk and the white
dwarf, on the surface of the white dwarf, or in a region where the winds from 
the two stars collide \citep[$\delta$, $\alpha$, and $\beta$ types, respectively]{Luna_etal_2013}, or the white dwarf fast wind collides with circumstellar material from an
earlier, slower wind \citep{Lucy_etal_2020}.

Radio surveys of SySts were carried out in the 1980s and 90s, mostly with the NRAO VLA telescope
\citep{Seaquist_etal_1984,Seaquist_Taylor_1990,Seaquist_Taylor_1992,Seaquist_etal_1993,Ivison_etal_1995}.
Radio fluxes are collected in table 1 of \citet{Belczynski_etal_2000}.  More recently, individual sources have been
monitored in the radio, particularly as target-of-opportunity projects to follow-up detections of 
outbursts in the optical and X-ray bands \citep{Brocksopp_etal_2004, Lindford_etal_2019}. In some cases the
radio flux is highly variable on time scales as short as a few weeks \citep[e.g.][]{Lucy_etal_2020}.  
There is good evidence from VLBI maps of transient radio jets that synchrotron emission is required, e.g. in R Aqr 
\citep{Dougherty_etal_1995}, in CH Cyg \citep{Crocker_etal_2001}, and others \citep{Rupen_etal_2008,Sokoloski_etal_2008,Giroletti_etal_2020}, to explain the occasional short-lived jet emission seen after some optical and X-ray outbursts. 
%The point of this sentence, as I read it, is that although there is evidence of highly variable radio activity, jets, synchrotron emission, etc, these things are not the norm. Jets are found after periods of high variability, a symbiotic can act nova-like and have emission that reflects those outbursts and produce synchrotron, but that's not the USUAL radio emission that we see. I think this is worth emphasizing, and I would re-write this sentence and move it to the end of the paragraph. :
%But apparently most of the radio emission is thermal bremsstrahlung most of 
%the time, as shown by the good correlation between radio luminosity and H$\beta$
% equivalent width \citep[fig. 4]{Seaquist_etal_1984}.  
The jets are typically found after optical outbursts 
\citep[and references therein]{Sokoloski_2003}.  Some SySts are recurrent novae, and SySt outbursts 
are similar in the radio to the emission from novae \citep[e.g.][]{Seaquist_Bode_2008, Weston_etal_2016}, particularly in their
mixture of rapidly brightening and fading non-thermal emission plus thermal emission with longer duration.
Apart from these short-duration synchrotron outbursts, apparently most of the radio emission from SySts is thermal
bremsstrahlung most of the time, as shown by the good correlation between radio luminosity and H$\beta$
 equivalent width \citep[fig. 4]{Seaquist_etal_1984}.  

One of the fundamental questions about SySts is whether or not there is shell burning, i.e. quasi-continuous
nuclear fusion, on the surface of the white dwarf \citep{Sokoloski_etal_2017}.  The radio luminosity can be
a diagnostic for the energy input due to shell burning.  Free-free (thermal) radio emission originates
in gas that is either 
photo-ionized by the white dwarf continuum \citep{Seaquist_etal_1984, Taylor_Seaquist_1984} or heated and ionized by
a strong shock in the colliding wind region.  
%The latter may power the hot gas that emits hard X-rays. 
If shell burning is happening on the white dwarf then both the ionizing radiation field
and the wind power will be much stronger than if it is not, so the radio luminosity of the system should
be stronger in systems with continuous nuclear fusion.  To test this hypothesis, 
\citet[][chap. 5]{Weston_2016} selected a sample of SySts that show characteristics
that indicate that they are powered by accretion and not by shell burning.  These
are sources that show UV flickering (rapid flux density variations caused by instabilities
on the inner edge of the accretion disk) and X-ray spectra indicating that the source of
the X-rays is either a colliding wind region [type $\beta$, \citet{Luna_etal_2013}]
or the boundary layer between the accretion disk and the white dwarf star (type $\delta$).
Many of these were detected in Weston's radio survey, but with flux densities one to two orders of
magnitude fainter than in surveys from 25 years earlier, that included all types of SySts.  

%[move this down to results?] In her thesis, \citet{Weston_2016}
%observed a sample of nearby SySts, selected based on the red giant distance modulus and no evidence for
%shell burning, to test whether there is a population of low radio luminosity systems.  She detected several
%at the level of a few tens of microJanskys, a factor of 10 to 100 fainter than most previous detections.
 
As part of a radio-infrared-optical-UV-X-ray simultaneous survey of X-ray binaries, we observed
several Southern Hemisphere SySts using the Australia Telescope Compact Array in June/July, 2007.  The detected flux
densities are similar to results of earlier studies \citep{Ivison_etal_1995}, but several of the sources
changed in flux density over the 10 or more years since they were observed before.  What is more useful
is the spectral indices of the sources.  Because the ATCA allows simultaneous observation in two bands
(C- and X-band, at 4.8 and 8.64 GHz) separated by almost a factor of two in frequency, the
spectral indices for our detections were measured with fairly good precision.  In this paper we discuss the
interpretation of the spectral index to determine the optical depth and brightness temperature of
the emission regions.  Since the SySts we observed now have \textcolor{cyan}{Gaia} %\textcolor{red}{
eDR3 distances \citep{Bailer-Jones_etal_2020},
%}
we can estimate the physical size as well as the solid angle of the ionized gas that emits the thermal
radio emission.  Comparing the results with similar quantities for the stars observed by
\citet{Weston_2016} shows that the systems selected not to have shell burning are
indeed very different from the rest.

%We also searched for variations in source flux densities over the seven days of our observations.  
%The brighter sources show no significant variations. 

\section{Radio Observations}\label{sec:Radio}
\begin{table}[!b]
\begin{center} 

%\caption{Compact Array Telescope Parameters }
\begin{tabular}{|l|ccc|} \hline 
E-W Baseline Lengths & 0.15 km & 0.41 km & 0.64 km\\
6C configuration& 1.06 km & 1.58 km & 1.73 km\\
& 1.99 km & 2.14 km & 2.63 km\\
& 2.79 km & 3.21 km & 3.86 km\\
& 4.27 km & 5.85 km & 6.00 km\\
\hline
Frequency Channels& 32 x 4 MHz &(24 used)&\\
\hline
\end{tabular}

\begin{tabular}{|l|c|c|}
\hline
band center& 4800 & 8640 MHz\\
\hline
resolution (FWHM)&  1.8\arcsec \ x 4\arcsec & 1.2\arcsec \ x 3.4\arcsec   \\
\hline
Primary beam width & 9.5\arcmin & 5.3\arcmin \\
\hline
Calibration Sources
 & S$_{4800}$ \textcolor{cyan}{(in Jy)} & S$_{8640}$ \textcolor{cyan}{(in Jy)} \\
0537-441  & 3.9 & 5.3 \\
0823-500&  3.1 &  1.5 \\
1059-63&  1.1  &  1.4\\
1352-63&  1.4 & 1.1 \\
1514-241 & 3.0 & 3.0 \\
1613-586  & 3.8 &3.3  \\
1730-130  & 4.1 & 3.6 \\
1740-517   & 4.1 & 2.6 \\
1741-038  & 5.8 & 5.7 \\
1759-39  & 1.3 & 1.5 \\
1921-293  & 9.5  & 10.9  \\
\hline
1934-638 & 5.83 & 2.84 \\
\hline
\end{tabular}
\caption{Compact Array Telescope Parameters.}
\label{tab:ATCA_params}
\end{center}
\end{table}

%\section{Radio Observations and Results}

%  tables:  tab:ATCA_params 
%           tab:positions (positions and flux or flux limit from UVFLUX), 
%           tab:radio_detections (position, peak and integ flux from mosaic maps)

The data were taken with the Australia Telescope Compact Array in the 6C
configuration, that has 15 East-West baselines as shown on Table \ref{tab:ATCA_params}.
The maximum projected baseline of 6 km is
1.7$\times$10$^5 \ \lambda$ at the highest observing frequency
of 8.64 GHz giving minimum fringe separation of 1.1\arcsec.  The observations
were taken over seven nights from 29 June - 5 July, 2007, in 12 hour
sessions from 10-22$^h$ LST. Two frequencies were
observed simultaneously, centered on 8.64 and 4.80 GHz.
% with two sessions centered 2.368 and 1.384 GHz on partial days July 2 and 3.  
The bandwidth was 128 MHz divided
into 32 channels on each of two polarizations, i.e.
correlator configuration full\_128\_2. The lowest 
%three 
%\textcolor{red}{
and%}
highest four
frequency channels are dropped to avoid filter edge effects, so the total
effective bandwidth is 
%100 
%\textcolor{red}{
96
%}
MHz at each frequency. 

The primary flux and bandpass calibrator was 1934-638, observed every day, with secondary
calibrators as shown on Table \ref{tab:ATCA_params}.  Secondary calibrators were
observed every 20 to 25 minutes,
interspersed with two scans on SySt program source positions observed for eight minutes each. 
Solutions for antenna phases and gains were interpolated between successive
calibrator scans and copied to the program source data, following the standard {\it MIRIAD}
calibration process
described in \citet{Sault_etal_1995}.  Calibrators were chosen
% Miriad Users Guide http://www.atnf.csiro.au/computing/software/miriad
within $\sim$20 degrees of the program source positions whenever possible.
Amplitude and phase errors were typically less than 5\% and 5\arcdeg \ as measured
on the calibrators at 8.64 GHz.  The weather was generally good with the exception
of the first few hours of 30 June, when tropospheric effects cause rapid, intermittent 
phase changes for several hours.  

Table \ref{tab:positions} gives phase center positions for each program source, and upper limit
flux densities for those that were not detected at 8.64 GHz. 
%\textcolor{red}{
These were the most accurate positions available at the time, but the positions
fitted to the cleaned maps are in some cases more precise.  So we 
mapped and cleaned the field of view for each source, fitted the source position, then
shifted the phase centers for all the {\it u,v} data to the new source position.  We 
%Although the {\it u,v} coverage was sparse on each day, we
used the {\it MIRIAD} task {\small UVFLUX}
to estimate the flux density on each day, based on the assumption of a point source
at the phase center.
%} 
The results are consistent with no flux variations over the
seven nights of observing.  We then put together all data to make maps, remove sidelobes with the 
{\small CLEAN} algorithm, and measure the position, flux density, and angular size 
%\textcolor{red}{
by fitting a two dimensional Gaussian source moodel
with task {\small IMFIT}.  
%}

%\textcolor{red}{
Table \ref{tab:radio_detections} gives the measured quantities for
each detected source at 8.64 and 4.80 GHz, based on the fitted Gaussian parameters:
peak brightness (column 3), integrated flux density (column 4), position (columns 7 and 8), and angular
diameter (column 9).  All the sources are unresolved, or barely resolved by the beam, which is generally
quite elliptical due to the sparse {\it{u,v}} coverage, typically $\sim$3.4\arcsec x 1.2\arcsec \ at 8.64 GHz.  The fitting program attempts to 
determine the source angular size, but this deconvolution fails if the fitted shape is
consistent with the beam shape, within the errors on the major and minor axes of the
fitted Gaussian.  In that case the upper limit major axis given on Table \ref{tab:radio_detections}
is the minor axis of the beam.
The integral of the fitted Gaussian, that gives the flux density, is computed from the product of
the fit results for the peak times the major and minor axes.  It necessarily has a 
larger error than the peak value, and it is usually somewhat larger than the
peak brightness, expressed as mJy/beam, due to the noise in the map, even for a
source that is unresolved.  
%}

%\textcolor{red}{
The conversion factor between mJy/beam
and K of brightness temperature is $\sim$3.5 K/mJy at 8.64 GHz and $\sim$5 K/mJy at 4.8 GHz,
depending on the beam size after tapering.  For example, ESO 393-31 has peak value
of $60\pm1.2$ mJy/beam at 8.64 GHz which gives peak brightness $T_B\simeq 210$ K.
The actual brightness temperature may be higher by a factor of five or more,
if the source solid angle is less than the beam solid angle, since the flux density is the 
brightness times the solid angle. In this case, IMFIT determines a source angular diameter of 0.9\arcsec.   
If the source solid angle, $\Omega = 1.13 \cdot (0.9)^2 \simeq 0.9$ arcsec$^2$ and
the beam solid angle is 4.6 arcsec$^2$ = 1.13 $\cdot$ 3.4\arcsec $\cdot$ 1.2\arcsec,
then the peak brightness is $T_B \sim 10^3$ K.  
(The factor of 1.13 comes from assuming Gaussian shape, for which the solid angle is
1.13 times the product of the major and minor axes, measured to the half-maximum points.)
The analysis of the spectral
index below in Section \ref{sec:analysis} predicts $T_B = 1.7\pm0.3 \cdot 10^3$ K assuming electron temperature
$T_e = 10^4$ K.  Reducing the assumed value of $T_e$ to $\sim 6 \cdot 10^3$ K reduces the 
predicted brightness temperature to $T_B \simeq 10^3$ K, which matches the measured 
surface brightness if the source solid angle is $\Omega \simeq0.9 (\textrm{arc\ sec})^2$.
%}

%\textcolor{red}{
Table \ref{tab:radio_detections} lists two components for NSV 19500 = SS73 38.  The 
brighter component is at the star position, while the second component, which may be
an unrelated source, is 53\arcsec \ NW of the star.  The probability of finding an
extragalactic source with flux density of 1 mJy or greater in a circle of radius
53\arcsec \ at 8.64 GHz is about 0.03 \citep{Condon_Ransom}.  At the eDR3 distance
of 2.19 kpc the distance on the plane of the sky corresponding to 53\arcsec \ is 
0.6 pc. 
%which makes it unlikely that the second source is related to NSV 19500.
%}

\begin{table}

\begin{center}

\begin{tabular}{|l|c|c|c|}\hline  
Source & RA & Dec & upper limit $S_X$ (mJy) \\
 \hline  
RX Pup &  8:14:12.31 & -41:42:29.0 & detected\\ 
KM Vel &  9:41:14.00 & -49:22:47.2 & detected\\
V366 Car &  9:54:43.29 & -57:18:53.0 & $<1.2$ \\
BI Cru &  12:23:25.99 & -62:38:16.1  & detected\\
NSV 19500 = SS73 38$^1$&  12:51:26.21 & -64:59:58.3 & detected\\ 
ESO 390-7 = Hen 2-171&  16:34:04.23$^2$ & -35:05:26.2$^2$ & $<1.0$\\ 
NSV 20790 = AS 210&  16:51:20.39 & -26:00:26.8 & detected\\ 
ESO 393-31 = PN H 1-36&  17:49:48.17 & -37:01:29.4 & detected\\
IRAS 18015-2709 = SS73 122&  18:04:41.28$^2$ & -27:09:13.6$^2$ & $<1.0$ \\ 
AN 68.1937 = V3929 Sgr &  18:20:58.85$^2$ & -26:48:25.1$^2$ & $<1.5$\\ 
RR Tel &  20:04:18.54 & -55:43:33.2& detected\\ 
 \hline
 \hline
\end{tabular} 
\caption{  Observed Field Center Positions \\
$^1$Alternate names are from table 1 of \citet{Belczynski_etal_2000}. \\
%\textcolor{red}{
$^2$Positions were taken from SIMBAD. There are offset from the positions
in \citet{Belczynski_etal_2000} of 8\arcsec for ESO 390-7, 19\arcsec for 
IRAS 18015-2709, and 7\arcsec for AN 68.1937. 
}%}
\label{tab:positions}

\end{center}
\end{table}
%\end{sidewaystable}

% position offsets:  ESO 390-7: 7.7"  our position agrees with "General Catalog of Variable Stars" and IRAS psc
%                                      PN G346.0+08.5 "Galactic Planetary Nebulae Catalog"
%                                     SIMBAD 16:34:04.23, -35:05:26.7
%                    IRAS 18015: 19.3"
%                                     SIMBAD 18:04:41.22, -27:09:12.4
%                    AN 68.1937: 6.9"
%                                     SIMBAD 18:20:58.87, -26:48:25.6

\begin{sidewaystable}[htp]
\centering{ 
%\vspace*{.2in}
%\textcolor{red}{ 
\begin{tabular}{|l|c|c|c|c|c|c|} 
\hline
 Source & Freq & peak brightness & flux density  & RA (peak)& Dec (peak) & Maj ax  \\
 & MHz & mJy/bm & mJy & J2000 & J2000 & $^{\prime\prime}$  \\
 \hline\hline
RX Pup & 8640 &24.4$\pm$0.6 & 30.0$\pm$1.1 &  8:14:12.30& -41:42:29.3 & 1.1 \\ 
RX Pup & 4800 & 25.3$\pm$0.5 & 27.9$\pm$0.8 &  8:14:12.31& -41:42:29.3 & 1.5   \\ 
 \hline
KM Vel & 8640 &3.3$\pm$0.1 & 3.3$\pm$0.3 & 9:41:14.00 & -49:22:47.2 & $<4$ \\ 
KM Vel & 4800 & 2.6$\pm$0.5 & 2.6$\pm$0.7 &  9:41:13.99 & -49:22:47.0 & $<7$ \\ 
 \hline
BI Cru & 8640 &4.0$\pm$0.2 & 4.3$\pm$0.3 &  12:23:25.97& -62:38:15.6& $<2$  \\ 
BI Cru & 4800 & 4.6$\pm$0.2 & 5.0 $\pm$0.4 &   12:23:25.98 & -62:38:15.6 & $<$3 \\ 
 \hline
NSV 19500 = SS73 38 &8640 & 7.4$\pm$0.2 & 7.4$\pm$0.5 & 12:51:26.17 &-64:59:57.7& $<$4  \\  
NSV 19500 = SS73 38& 4800 & 5.2$\pm$0.3 & 5.0$\pm$0.4 & 12:51:26.19 &-64:59:57.8& $<$4  \\  
Second source component: &&&&&&\\
NSV 19500 b &8640 & 1.1$\pm$0.2 & 1.1$\pm$0.3 & 12:51:18.16 &-64:59:53.9& $<$4  \\  
NSV 19500 b &4800 & 1.4$\pm$0.2 & 1.5$\pm$0.3 & 12:51:18.14 &-64:59:53.9& $<$4  \\  
 \hline
NSV 20790 = AS 210&8640 &5.6$\pm$0.3 & 5.7$\pm$0.4 & 16:51:20.40 &-26:00:26.7 & $<3$   \\ 
NSV 20790 = AS 210&4800 &4.9$\pm$0.5  &5.7$\pm$0.9  &16:51:20.40 &-26:00:26.6 &$<$3   \\
 \hline
ESO 393-31 = PN H 1-36& 8640 &60.0$\pm$1.2 & 65.5$\pm$1.8& 17:49:48.21& -37:01:27.8& 0.9  \\ 
ESO 393-31  = PN H 1-36& 4800 & 53.2$\pm$0.7 & 56.1$\pm$1.0 & 17:49:48.21 & -37:01:27.8 & 0.9\\ 
 \hline
RR Tel & 8640 &12.8$\pm$1.2 & 16.2$\pm$2.3 & 20:04:18.54 & -55:43:32.9 & $<$2  \\ 
RR Tel & 4800 & 15.3$\pm$0.5 & 17.6$\pm$0.8 &  20:04:18.54 & -55:43:33.0 & 1.3   \\ 
 \hline
 \hline
\end{tabular}}
\caption{Detected Source Properties}   \label{tab:radio_detections} 
%}
\end{sidewaystable}

\section{Analysis} \label{sec:analysis}

Recent hardware upgrades to the VLA and ATCA telescopes provide %\textcolor{red}{
increased%}
bandwidth which results in higher sensitivity \citep{Perley_etal_2011,CABB}. Even the older
receiver system on the ATCA used here was versatile in allowing observation 
with two bands simultaneously.  Using the CX system with the new CABB 
receivers now allows many bands to be placed between about 4 and 10 GHz.
This makes measurement of the spectral index, $\alpha$, easier and more
direct than in the original single-band VLA system that was used for the
comprehensive surveys by Seaquist and collaborators, summarized by 
\citet{Ivison_etal_1995}.  The spectral index is important for interpretation
of the radio continuum data, particularly thermal emission like that from
most SySts.  The discussion in this section follows the excellent textbook review
of thermal radio emission by \citet[chap. 4]{Condon_Ransom}.

The spectral index, $\alpha$, can be defined at a single frequency as the slope
of the flux density spectrum in log-log space, e.g. \citet[figure 6]{Seaquist_etal_1984}, but it is more
commonly measured between
%\textcolor{red}{
two frequencies, $\nu_1$ and $\nu_2$ using the corresponding flux densities
$S_{1}$ and $S_{2}$, by 
\begin{equation} \label{eq:alpha_def}
 \alpha = \frac{\log{\left(\frac{S_{1}}{S_{2}}\right)}}{\log{\left(
\frac{\nu_1}{\nu_2}\right)}} .
\end{equation}
%In Equation \ref{eq:alpha_def} the logarithms can have any base, but in eq. \ref{eq:sigma_alpha_1} below
%there is specifically a natural log (ln), and on Table \ref{tab:results} and
%Figure \ref{fig:alpha_4panel} logs base 10 are indicated by "log".
%}
With this definition thermal emission has $\alpha$ = +2, for the
Rayleigh-Jeans (optically thick) black-body spectrum, decreasing to
$\alpha$ = -0.1 for optically thin free-free emission.  In the simplest case of a uniform
slab of ionized gas with
electron temperature, $T_e$, and density, $n_e$, the optical depth, $\tau$,
depends simply on frequency, $\nu$, to the -2.1 power, as
\begin{equation}\label{eq:tau} \tau_{\nu} \ = \ C_1 \ \nu^{-2.1} \end{equation}
where the constant $C_1$ is the same at frequencies $\nu_1$ and $\nu_2$
as long as the emission at both frequencies comes from the same gas.
This uniform slab assumption is obviously an over-simplification, but it allows
us to derive representative numbers for the physical parameters by simple algebra,
that otherwise might be buried in a more realistic gas dynamic and
radiative transfer model.  $C_1$ is given by
\begin{equation} \label{eq:C1}
%\textcolor{red}{
 \frac{C_1}{\textrm{(GHz)}^{2.1}} \ = 
\ 3.28 \cdot 10^{-7} \  \int \left(\frac{n_e}{\textrm{cm}^{-3}}\right)^2 \ \left(\frac{T_e}{10^4 \textrm{\ K}}\right)^{-1.35} \ \frac{ds}{\textrm{pc}} \ \simeq 
 \ 3.28\cdot 10^{-7} \  \left(\frac{T_e}{10^4 \ \textrm{K}}\right)^{-1.35} \ \frac{EM}{\textrm{cm}^{-6} \ \textrm{pc}} %} 
\end{equation}
where the integral is taken along the line of sight, $s$,
and the emission measure, $EM$, is the line of sight integral of the density
squared:
$EM \ = \ \int n_e^2 \ ds $.
If the emission at the two frequencies is from the same volume, then the 
spectral index is given by
%\textcolor{red}{
\begin{equation}\label{eq:alpha_formula} 
\alpha = \ \frac
{ \log{\left[ \left(\frac{\nu_1}{\nu_2}\right)^2  \  \cdot \ 
\frac{ 1\ - \ exp({- C_1 \ \nu_1^{-2.1}})}{ 1\ - \ exp({- C_1 \ \nu_2^{-2.1}})} 
\right] }} 
{\log{\left(\frac{\nu_1}{\nu_2}\right)}} \end{equation}
%}
For a given pair of frequencies and the corresponding value of $\alpha$,
we can solve numerically for the value of $C_1$, giving the optical depth 
and hence the ratio of the brightness temperature, $T_B$, to $T_e$, since
\begin{equation} \label{eq:TB}
T_{B,\nu} \ = \ T_e \ \left(1 \ - \ e^{-\tau_{\nu}} \right) \end{equation}
where $T_B$ and $\tau$ are functions of $\nu$, easy to evaluate once $C_1$ is known.

%\textcolor{red}{
For different geometries the constants vary, e.g.  
\citet{Mezger_Henderson_1967},
\citet{Seaquist_etal_1984}, and see the two component model in section \ref{sec:appendix}.
The radiative transfer can be modeled for time-varying conditions, e.g. for an expanding sphere as in 
\citet{Wright_Barlow_1975} and \citet{Seaquist_Palimaka_1977}.
%\citet{Contini_etal_2009},
For the sake of simplicity, we continue with the uniform slab model 
to estimate representative values for the properties of the
emission region based only on $S_{1}$ and $\alpha$.  If the ionized region  %}
has a sharp edge and uniform brightness temperature, then the 
flux density is 
\begin{equation}\label{eq:Snu}
 S_{\nu}\ = \ \frac{2 \ k \ T_B \ \nu^2}{c^2} \ \Omega \ = \ 
\frac{T_B}{10^4 \ \textrm{K}} \  \ \frac{\Omega}{(\textrm{arc \ sec})^2} \ 
\cdot \ \left\{
\begin{array}{ll} 539 \ \textrm{mJy} & \textrm{at}\ \nu = 8.64\ \textrm{GHz}\\
585 \ \textrm{mJy} & \textrm{at} \ \nu = 9.0 \ \textrm{GHz} \end{array}\right. \end{equation}
where $k$ is Boltzmann's constant and $c$ the speed of light.
For a resolved source the solid angle, $\Omega$, can be determined by deconvolving the beam
shape, and the result depends on the size of the source; if the source is
much smaller than the beam, then neither $T_B$ nor $\Omega$ can be measured
directly, only their product, i.e. $T_B \Omega = S_{\nu}\ c^2$/($2 \ k\  \nu^2$).  
% This is WRAY-15 1470. It's interesting that this one is the outlier -- I discussed this in chapter 5 of my thesis, but this was source was never observed to have delta or beta/delta type X-ray emission, and had the lowest measured UV flux ratios in the survey; it was chosen simply because it had high UV variability. So this was not as certain to be accretion driven as some of the other sources. I wonder if it might be worth addressing this somewhere (a footnote? A reference to that statement in my thesis?), because it does stand out a bit on the plot! If you want me to draft up some language on that, let me know. I promise to be a lot more prompt this time!!! :)

\begin{figure}[h]
\includegraphics[width=6.5in]{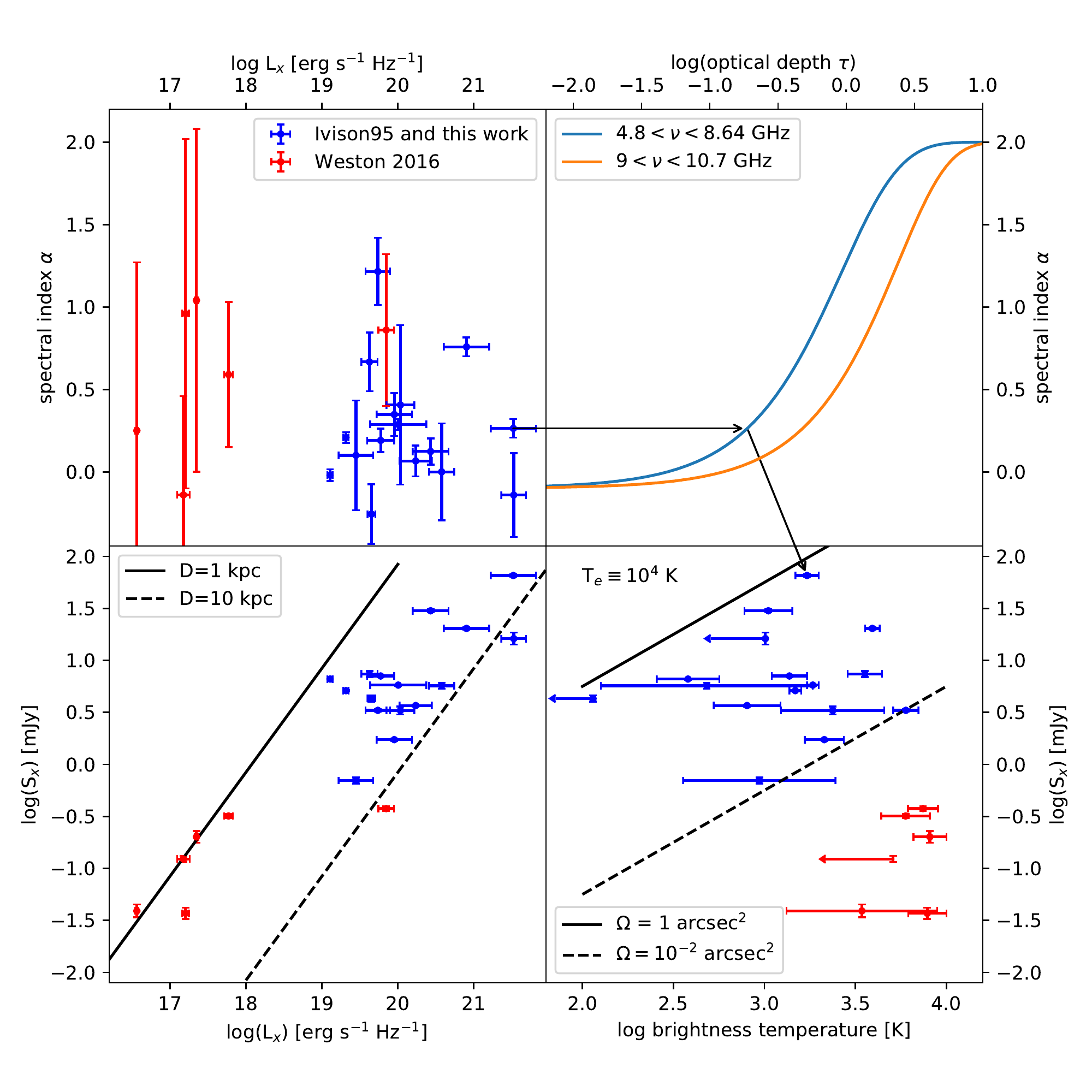}
\caption{Spectral index analysis of radio detections.  The lower left panel shows the
distribution of flux densities at X-band, $S_X = S_{\nu}$ where $\nu$ = 8.64 GHz for
the values from \citet{Ivison_etal_1995} and this work, or
$\nu$=9.0 GHz for the values from \citet{Weston_2016}, plotted vs. 
%\textcolor{red}{
monochromatic %}
luminosity,
$L_{\nu}$ in units of erg s$^{-1}$ Hz$^{-1}$.  The luminosities are based on
distances from Gaia
eDR3 tabulated by \citet{Bailer-Jones_etal_2020}.%} 
The upper left panel shows
the corresponding values of the spectral index, $\alpha$ for each source.  Based
on the frequencies used to measure $\alpha$, we find the corresponding value of optical depth
from the curves on the upper right panel (equation \ref{eq:alpha_formula}).  With the optical
depth known we find the brightness temperatures shown in the lower right panel if we assume
a value for the electron temperature, in this case $T_e$=10$^4$ K.  Choosing a different value for
$T_e$ rescales the x-axis %\textcolor{red}{
of the lower right panel%}
by a factor of $\frac{T_e}{10^4 \  K}$, and it scales the solid angles by the inverse, i.e. $\left(\frac{T_e}{10^4 \ K}\right)^{-1}$ (see section \ref{sec:appendix}),
since the flux density is the product of the brightness temperature times the solid angle in this simple geometry. The black arrows illustrate this strategy for one case %\textcolor{red}{
(ESO 393-31).  All points have error bars in both directions, but they are not plotted if the
errors are smaller than the symbols.  %}
}
\label{fig:alpha_4panel}
\end{figure}

Figure \ref{fig:alpha_4panel} illustrates the derivation of brightness temperature and solid angle (right hand panels)
based on the measured value of $\alpha$ (upper left panel).

Precision in the measurement of $\alpha$ is critical.  Propagating errors
in equation \ref{eq:alpha_def}  gives dispersion $\sigma_{\alpha}^2$ where
%\textcolor{red}{
\begin{equation} \label{eq:sigma_alpha_1}
\sigma_{\alpha} \ = \ \frac{S_{2}}{S_{1}} \ \frac{1}{\left|\ln{\left(
\frac{\nu_2}{\nu_1}\right)}\right|} \ \sqrt{
\frac{\sigma_{S_1}^2}{S_2^2} \ + \ \frac{S_1^2}{S_2^4} \ \sigma_{S_2}^2\ } \end{equation}
%}
for an estimate of $\alpha$ based on measurements of $S_1$ and $S_2$ with their own
dispersions of %\textcolor{red}{
$\sigma_{S_1}^2$ and $\sigma_{S_2}^2$.  %}
% ref If the errors in the measurements
% ref of the two flux densities are about the same, i.e. $\sigma_{S_1} \simeq \sigma_{S_2} \simeq \sigma_{S}$, then
% ref \begin{equation}\label{eq:sigma_alpha_2}
% ref \sigma_{\alpha} \ \simeq \ \frac{\sigma_{S}}{S_{1}} \ \frac{1}{\log{\left(\frac{\nu_2}{\nu_1}
% ref \right)} } \ \sqrt{1 \ + \ \left(\frac{S_1}{S_2}\right)^2 \ }
% ref \end{equation}
The logarithmic term in the denominator shows that the two frequencies need to 
be separated by as large a factor as possible.  
Even with 10:1 signal to noise, i.e. $\frac{\sigma_S}{S} \sim 0.1$, the precision
in $\alpha$ is marginal, $\sigma_{\alpha}\simeq 0.6$,
for $\nu_1$=4.8 GHz and $\nu_2=8.64$ GHz. If the frequency range is just 9.0 to 10.7 GHz
as for the X-band system on the Karl Jansky Very Large Array
then the error in $\alpha$ increases to 1.9.  
One of the advantages of the ATCA
for this work is that its CX band allows a wide range of frequencies to be observed
simultaneously.  

In general several different sources of error contribute to the measurement
of flux density, $S$, which complicates the determination of $\sigma_{\alpha}$.
Ordinary radiometer noise adds random numbers from a well-determined,
Gaussian distribution that can usually be attenuated with longer integration.  
Random scale factors multiply the measured values by random numbers 
with a distribution that depends on the weather, the stability of the system,
the observing strategy, and the level of flagging, i.e. rejection of data
of mediocre quality.  Generally the precision of flux densities measured at
X-band ($\lambda \simeq $ 3.5 cm) is 3\% to 5\%.  If there are other radio 
sources within the primary beam then confusion can dominate the
error in $S$; if the {\it u,v} coverage is sparse then the flux from the other
sources will spread around the image plane as noise-like fluctuations that
limit the dynamic range of the image.  In this experiment the
primary beam is small enough that there are no bright confusing sources in the 
fields of the SySts at 4.8 and 8.64 GHz, but gain errors associated
with weather are noticeable.
For the source ESO 393-31 = PN H 1-36 these dominate the errors, but for
the other sources the flux densities are small enough that radiometer noise
dominates.  So we use the simple formula of equation \ref{eq:sigma_alpha_1} to determine 
$\sigma_{\alpha}$, noting that these numbers may be underestimates.

Many SySts show parallax in the Gaia Data Release %\textcolor{red}{
3 (eDR3), \citet{Bailer-Jones_etal_2020}.%}.
% Akras, S., Guzman-Ramirez, L., Leal=Ferreira, M.L., and Ramos_larios, G., 2019, \apjs 240, 21A.
Although the resulting distance estimates come with fairly large error-bars for
the more distant objects, the %\textcolor{red}{
Gaia values
%}
are much more reliable than previous
distance estimates, that were mostly based on photometry of the red giant star.
Now we can convert flux to luminosity, and solid angle to projected area.
Table \ref{tab:results} gives the resulting values, for the sample of stars observed in this
study, and for those with eDR3 distances in the collection of
\citet{Ivison_etal_1995} that have
flux densities measured at $\lambda\lambda$ 3.5 and 6.25 cm, and for those with eDR3 distances in the
survey of \citet{Weston_2016}.
%Included on Table \ref{tab:results} are objects with estimates for $\alpha$ from this work
%or the surveys tabulated by \citet{Ivison_etal_1995} that do not have eDR3 distances in \citet{Bailer-Jones_etal_2020} (indicated by dots in columns 3, 4, 10, and 11).
%(FWHM), giving $\Omega \simeq$ (FWHM), giving $\Omega \simeq$ 0.4 (arc sec)$^2$.

Column 1 of Table \ref{tab:results} gives the source name, column 2 gives the X-band
flux density $S_{\nu}$ either at 
$\nu$=8.64 GHz [from this work or \citet{Ivison_etal_1995}] or at $\nu$=9.0 GHz [from \citet{Weston_2016}].
Column 3 shows the three eDR3 values: minimum, best, and maximum distance ($D$), in pc.  We use the middle value for
$D$ to find the luminosity and source size.  
Column 4 gives the resulting luminosity, $L_X = 4 \pi (D)^2 S_X$.  Column 5 gives the measured value of
$\alpha$, with error, and column 6 gives the resulting value of $\tau$ at 8.64 GHz, or 9.0 GHz for the \citet{Weston_2016} sources.
The values in the next five columns depend on the electron temperature in the
emission region, here we take $T_e$ = 10$^4$ K, but see section \ref{sec:appendix} for a discussion of how these change with changing $T_e$.
Column 7 gives the log$_{10}$ of the emission measure, $EM$, implied by $\tau$.   
%For the stars with distances 
%\textcolor{red}{
Using the Gaia distances%}
we can determine the linear size from the solid angle, assuming
some source geometry.  We can get a characteristic number for the diameter, $d  =  (D) \sqrt{\frac{4}{\pi} \Omega }$ by assuming a simple cylinder of hot gas.
The sizes, $d$, are given on column 10 of Table \ref{tab:results}.  If we assume that
the line-of-sight depth is the same as $d$, then the electron density is given by 
$n_e = \sqrt{EM/d}$, given in column 11.  For different values of the assumed electron temperature,
$d$ scales as 1/$\sqrt{(T_e/10^4)}$ and $n_e$ scales as ($T_e/10^4)^{0.25}$.

The brightest source detected in the 2007 observations is ESO 393-31 = PN H 1-36.  It is classified
as a SySt, but the radio emission is probably dominated by the extended ionized region of the 
associated planetary nebula.  %\textcolor{red}{
The prediction for $\Omega \simeq 0.8$ (arc sec)$^2$ is larger for this source than
any of the others on Table \ref{tab:results} that have measured values of $\tau$.  (Three sources
have only upper limits for $\tau$, hence upper limits on $T_B$ and lower limits on $\Omega$.)  As noted above,
this source is marginally resolved, with deconvolved angular size $\sim$0.9\arcsec, giving
$\Omega \simeq$ 0.9 (arc sec)$^2$, in good agreement with the prediction.%}
All the other derived values of $\Omega$ on Table \ref{tab:results} suggest angular diameters
between 0.1\arcsec \xspace and 1\arcsec \xspace for the first group, i.e. those with flux densities $S_X$ of a few mJy.
%\textcolor{red}{
At the eDR3 distances the linear diameters are a few hundred to a few thousand AU.%}
The second group, those observed by \citet{Weston_2016} with flux densities of a few to a few
tens of $\mu$Jy, shows four sources, UV Aur, ZZ CMi, T Crb and ER Del, that have much smaller sizes 
with implied diameters $\sim$ 10 to 50 AU.  
Assuming that the power required to maintain the temperature and
ionization in the emission regions is proportional to the collision rate and the total volume of ionized gas, i.e. $n_e^2 \cdot r^3$, then the first group
of sources require about 100 times as much power as the four smaller, higher density sources in
the second group based on the diameters and densities in columns 10 and 11 on 
Table \ref{tab:results}.

The enigmatic system WRAY 151470, observed by \citet{Weston_2016} and listed on the last line on Table \ref{tab:results}, is unlike
either the high luminosity sources in the first group, or the low luminosity sources in the second group.  On the left panels of
Figure \ref{fig:alpha_4panel} this source stands out as the red point among the blue points on the right hand (high luminosity) side.
This SySt was selected because of its high UV variability \citep{Luna_etal_2013}, but it does not have a $\delta$ or 
$\beta$/$\delta$ X-ray spectrum, nor does it have a high UV flux ratio \citep[pp. 148-149]{Weston_2016}.  
%\textcolor{red}{
\citet[pp. 157-158]{Weston_2016} notes that WRAY 151470 therefore shows somewhat weaker evidence for being a purely accretion driven system than the other targets of that survey, and this anomalous behavior might be explained if there were some 
degree of shell burning present or if it were in a particularly active state.
Based on its eDR3 distance, it now appears to fit better in the class of
SySts with high radio luminosity, i.e. those above the line on Table \ref{tab:results}.%}

The striking difference between the indicated sizes and densities of the
ionized regions in the two groups of SySts supports the interpretation of \citet{Weston_2016} and
\citet{Sokoloski_etal_2017} that the radio luminosity can distinguish between stars with 
shell burning and those without.  
Looking at the lower left panel of Figure \ref{fig:alpha_4panel} the wide range of radio luminosities,
$L_{\nu}$, is clear.  None of the existing radio surveys has used a volume-limited sample,
so we cannot begin to draw a luminosity function for the radio emission from SySts.  But 
the distribution of points on the figure is suggestive of a bimodal luminosity function, with
%peaks around 10$^{17}$ and 10$^{20}$ erg cm$^{-2}$ s$^{-1}$ Hz$^{-1}$.  Interpreting these as
peaks around 10$^{17}$ and 10$^{20}$ erg s$^{-1}$ Hz$^{-1}$.  Interpreting these as
the off- and on-states of shell burning in the SySts, these luminosity values may be useful for
modelling the energetics of the ionized regions.

%\end{document}

.%\documentclass[10pt,preprint]{aastex1
%\usepackage{rotating}
%\pagestyle{empty}
%\begin{document}
\begin{sidewaystable}[ht]
\centering
%\textcolor{red}{
\begin{tabular}{|l|c|c|c|c|c|c|c|c|c|c|} 
\hline
 Source & S$_X$ & eDR3 D (kpc) & log(L$_X$)  & $\alpha_X$ & $\tau_X$ & log(EM) & T$_{B,X}$  & log($\Omega$) & diameter & log(n$_e$) \\ 
  & mJy & min best max & erg s$^{-1}$ Hz$^{-1}$  &  &  & cm$^{-6}$ pc & 10$^3$ K  & (arcsec)$^2$ & 10$^2$ AU & cm$^{-3}$ \\ 
&&&&&&\multicolumn{5}{c|}{ --- --- --- \hspace{\stretch{1}} estimates assuming T$_e$=10$^4$ K \hspace{\stretch{1}} --- --- ---} \\ 
\hline
RX Pup & 30.0$\pm$1.1 & 2.22  2.74  3.60 & 20.43$\pm$0.24 & 0.12$\pm$0.08 & 0.11$\pm$0.04 &  7.50$\pm$0.14 &  1.1$\pm$0.4 &  -0.28$\pm$0.13 & 22$\pm$3 &   4.7$\pm$0.1 \\
KM Vel &  3.3$\pm$0.3 & 4.33  5.24  6.47 & 20.03$\pm$0.18 & 0.4$\pm$0.5 & 0.3$\pm$0.3 &  7.9$\pm$0.3 &  2.4$\pm$2.2 &  -1.6$\pm$0.3 &  9$\pm$3 &   5.1$\pm$0.3 \\
BI Cru &  4.3$\pm$0.3 & 2.75  2.96  3.14 & 19.65$\pm$0.05 & -0.26$\pm$0.18 & $<\ $ 0.01 & $<\ $ 6.5 &  $<\ $ 0.1 &   $>\ $ -0.2 & $>\ $ 28 &   $<\ $ 4.2 \\
NSV 19500 &  7.4$\pm$0.5 & 1.91  2.19  2.47 & 19.63$\pm$0.11 & 0.67$\pm$0.18 & 0.44$\pm$0.14 &  8.09$\pm$0.12 &  3.6$\pm$0.9 &  -1.41$\pm$0.09 &  4.8$\pm$0.5 &   5.4$\pm$0.1 \\
NSV 20790 &  5.7$\pm$0.4 & 6.10  7.44  9.00 & 20.58$\pm$0.17 & 0.0$\pm$0.3 & 0.05$\pm$0.15 &  7.1$\pm$0.6 &  0.5$\pm$1.4 &  -0.7$\pm$0.6 & 39$\pm$19 &   4.4$\pm$0.5 \\
ESO 393-31 & 65.5$\pm$1.8 & 4.72  6.50  9.15 & 21.5$\pm$0.3 & 0.26$\pm$0.06 & 0.19$\pm$0.03 &  7.73$\pm$0.07 &  1.7$\pm$0.3 &  -0.15$\pm$0.06 & 62$\pm$4 &   4.62$\pm$0.05 \\
RR Tel & 16.2$\pm$2.3 & 10.6  13.2  15.9 & 21.53$\pm$0.16 & -0.14$\pm$0.25& $<$ 0.11 &  $<$ 7.5 &  $<$ 1.0 & $>$ -0.5 & $>$ 81  &  $<$ 4.4  \\
V835Cen &  5.8$\pm$0.1 & 2.78  3.83  5.86 & 20.0$\pm$0.4 & 0.29$\pm$0.03 & 0.20$\pm$0.02 &  7.76$\pm$0.04 &  1.8$\pm$0.1 &  -1.24$\pm$0.03 & 10.4$\pm$0.4 &   5.03$\pm$0.03 \\
He 2-127 &  0.7$\pm$0.1 & 4.16  5.79  7.54 & 19.45$\pm$0.23 & 0.1$\pm$0.3 & 0.10$\pm$0.18 &  7.4$\pm$0.5 &  0.9$\pm$1.5 &  -1.9$\pm$0.4 &  7.7$\pm$2.9 &   4.9$\pm$0.3 \\
AG Peg &  6.6$\pm$0.1 & 1.22  1.27  1.32 & 19.11$\pm$0.03 & -0.02$\pm$0.04 & 0.04$\pm$0.02 &  7.0$\pm$0.2 &  0.4$\pm$0.2 &  -0.49$\pm$0.17 &  8.2$\pm$1.5 &   4.7$\pm$0.1 \\
He 2-38 &  1.7$\pm$0.1 & 5.15  6.59  8.61 & 19.95$\pm$0.23 & 0.35$\pm$0.13 & 0.24$\pm$0.08 &  7.83$\pm$0.12 &  2.1$\pm$0.6 &  -1.82$\pm$0.11 &  9.1$\pm$1.1 &   5.1$\pm$0.1\\
SS 38 &  5.1$\pm$0.1 & 1.77  1.84  1.92 & 19.32$\pm$0.04 & 0.21$\pm$0.03 & 0.16$\pm$0.01 &  7.65$\pm$0.04 &  1.5$\pm$0.1 &  -1.19$\pm$0.03 &  5.3$\pm$0.2 &   5.12$\pm$0.03 \\
He 2-104 &  3.3$\pm$0.1 & 3.10  3.71  4.48 & 19.74$\pm$0.16 & 1.2$\pm$0.2 & 0.9$\pm$0.3 &  8.41$\pm$0.12 &  6.0$\pm$1.0 &  -1.99$\pm$0.07 &  4.2$\pm$0.3 &   5.5$\pm$0.1 \\
HD149427 & 20.3$\pm$0.5 & 3.69  5.75  8.10 & 20.9$\pm$0.3 & 0.76$\pm$0.06 & 0.50$\pm$0.06 &  8.15$\pm$0.05 &  3.9$\pm$0.4 &  -1.02$\pm$0.04 & 20.1$\pm$0.9 &   5.08$\pm$0.04 \\
He 2-176 &  3.7$\pm$0.1 & 4.79  6.25  7.98 & 20.23$\pm$0.21 & 0.07$\pm$0.09 & 0.08$\pm$0.05 &  7.4$\pm$0.2 &  0.8$\pm$0.4 &  -1.1$\pm$0.2 & 21$\pm$4 &   4.7$\pm$0.1 \\
K3-9 &  7.1$\pm$0.2 & 2.31  2.65  3.25 & 19.77$\pm$0.18 & 0.19$\pm$0.07 & 0.15$\pm$0.04 &  7.62$\pm$0.11 &  1.4$\pm$0.3 &  -1.02$\pm$0.10 &  9.3$\pm$1.0 &   4.98$\pm$0.08 \\
\hline
UV Aur &  0.20$\pm$0.03 & 0.94  0.96  0.99 & 17.35$\pm$0.02 & 1.0$\pm$1.0 & 1.7$\pm$8.2 &  8.7$\pm$0.8 &  8.1$\pm$1.9 &  -3.37$\pm$0.09 &  0.22$\pm$0.02 &   6.3$\pm$0.3 \\
ZZ CMi &  0.32$\pm$0.02 & 1.18  1.25  1.33 & 17.77$\pm$0.06 & 0.6$\pm$0.4 & 0.9$\pm$0.8 &  8.4$\pm$0.3 &  6.0$\pm$2.1 &  -3.04$\pm$0.13 &  0.42$\pm$0.06 &   6.1$\pm$0.4 \\
NQ Gem &  0.12$\pm$0.01 & 0.94  1.01  1.11 & 17.18$\pm$0.08 & -0.1$\pm$0.6 & $<$ 0.7 & $<$ 8.3  &  $<$ 5 & $>$ -3.4  &  $>$ 0.23  &  $<$ 6.1   \\
T CrB &  0.04$\pm$0.01 & 0.86  0.89  0.91 & 16.56$\pm$0.02 & 0.3$\pm$1 & 0.4$\pm$1.8 &  8.1$\pm$0.7 &  3.4$\pm$5.5 &  -3.7$\pm$0.4 &  0.14$\pm$0.05 &   6.1$\pm$0.2 \\
ER Del &  0.04$\pm$0.01 & 1.81  1.90  2.01 & 17.20$\pm$0.05 & 1.0$\pm$1.1 & 1.5$\pm$8.4 &  8.7$\pm$0.8 &  7.9$\pm$2.1 &  -4.09$\pm$0.10 &  0.19$\pm$0.02 &   6.4$\pm$0.3 \\
WRAY151470 &  0.38$\pm$0.02 & 10.7  12.5  14.1 & 19.85$\pm$0.10 & 0.9$\pm$0.5 & 1.4$\pm$0.9 &  8.6$\pm$0.2 &  7.4$\pm$1.6 &  -3.06$\pm$0.08 &  4.1$\pm$0.4 &   5.7$\pm$0.3 \\
\hline
 \end{tabular}
%}
\caption{Results \textcolor{cyan}{for the physical parameters of the emission regions based on the spectral index analysis described in equations 2 - 7.}} \label{tab:results} \end{sidewaystable}
% \end{document}

\section{Future Observations} \label{sec:future}

The results of the analysis above show the importance of accurate measurement of the 
spectral index, $\alpha$, for the interpretation of radio emission from SySts.
A more sophisticated approach would be to measure the fluxes over a range
of frequencies, if possible covering a factor of 3 to 5 in $\nu$ centered near 8 GHz,
where most of the sources are becoming optically thin.
This may require switching receivers at short intervals so that the {\it{u,v}} coverage
is nearly identical, in case of source structure or confusing sources in the primary beam.  Repeat observations in different telescope configurations will be needed if the emission is strongly resolved.
Given four or five accurate measurements of $S_{\nu}$ from about 4 to 16 GHz, spectral
indices and optical depths could be determined %\textcolor{red}{
with good precision.%} 
If the conditions in the emission
region are not uniform, then the shape of the radio spectrum in this frequency range could
allow separation of components with different values of $\tau$ (section \ref{sec:appendix}).

One of the most interesting parameters to measure is the electron temperature, $T_e$, in
the emission region, that is used in this analysis to translate from optical depth
to brightness temperature.
On Table \ref{tab:results} we use $T_e = 10^4$ K which is somewhat
higher than in most HII regions, but lower than
in planetary nebulae \citep{Zhang_etal_2004}.  It is possible that some of the ionized gas
seen in the radio is much hotter, see section \ref{sec:appendix}.  The X-ray emission is
thermal in all well studied SySts, with implied temperatures as
high as 10$^7$ K $\sim$ 1 keV (for $\beta$-type systems, see \citet{Luna_etal_2013}).
The white dwarf stars have surface temperatures in the range 2$\cdot10^4$ to
3$\cdot10^5$ K \citep{Contini_etal_2009}.  The temperature in the colliding wind region
of AG Peg has been determined spectroscopically by \citet{Nussbaumer_etal_1995} as $T_e \simeq 2\cdot10^4$ K
and density in the range $10^7$ to $10^{10}$ cm$^{-3}$, much higher than the densities implied
by the radio emission assuming $T_e$ = 10$^4$ K, (Table \ref{tab:results}, column 11).
The cleanest way to determine the electron temperature in the radio emission region is
to use the line-to-continuum ratios of H recombination lines with quantum levels $n \sim 80$ to 100
as has been done for HII regions \citep{Brown_etal_2017,Wenger_etal_2019}.  This takes long integrations even for
sources with $S_X \sim 100$ mJy.  For measurements of radio recombination lines in SySts we may have to wait for the
Next Generation VLA telescope %\textcolor{red}{
\citep{Balser_etal_2018} and the Square Kilometre Array \citep{OBrien_etal_2015}.%}.

Direct measurement of the sizes of the emission regions in more SySts would be very helpful.
VLBI studies such as the recent combined EVN-VLBA monitoring of V 407 Cyg by \citet{Giroletti_etal_2020}
have surface brightness sensitivity limited to $T_B > 10^5$ K, so they mostly show the distribution
of non-thermal emission, although if some of the radio flux comes from the X-ray emission region
it might be remotely possible that in a few very high density clumps the brightness temperature could be high enough to detect their thermal
emission with VLBI.  But thermal emission from regions with sizes on the order of 10$^{-1}$ arc sec could be measured
at higher frequencies with ALMA or the JVLA, or possibly with the ATCA.

Monitoring $S_X$ from a number of SySts of both high and low radio luminosities with a cadence of
a few weeks to months with photometric precision of 5\% or better would be of interest.
%Since the emission region sizes derived on Table \ref{tab:results} correspond to a light-hours to
%light-days, it is possible that rapid variations in the white dwarf luminosity could generate
%variations in the radio luminosity.
The recombination time for gas at $T_e\sim 10^4$ K
and $n_e \sim 10^5$ cm$^{-3}$ is about 460 days, for $n_e \sim 10^6$ cm$^{-3}$ it is 46 days
\citep[][case B, fig 14.1]{Draine_2011}, thus the lower luminosity sources would be most likely to show 
flux variations on time scales of a few weeks to months.

%\vspace*{0.5in}

\vspace*{.2in}

\section{Conclusions}\label{sec:Conclusions}

Based on the numbers in column 4 of Table \ref{tab:results} and the lower left panel of
Figure \ref{fig:alpha_4panel}, a dividing line between burning and
non-burning white dwarf stars in symbiotic binaries is given by X-band (8.6 GHz) luminosity $\log_{10}(L_X) \sim 18.5$.
SySts with %\textcolor{red}{
monochromatic luminosity (in erg s$^{-1}$%}
 Hz$^{-1}$) above 10$^{18.5}$
can ionize an HII region with size of 400 to 6000 AU.  These ionized regions almost qualify as ultracompact
HII regions \citep[Table 2]{Churchwell_1990}, that typically have emission measures between 10$^7$ and 10$^9$ 
pc cm$^{-6}$ and sizes $\sim 5 \cdot 10^{-3}$ up to $\sim$0.1 pc.  Unlike ultracompact HII regions, the radio
bright SySts are buried in a red giant wind rather than a molecular cloud, and they may not be entirely photoionized
but instead or in addition they may be heated and ionized by a colliding-wind shock.  The low radio luminosity SySts (UV Aur, ZZ CMi,
T Crb and ER Del) have higher emission measures, but much smaller implied diameters, than the high radio luminosity systems.  Either the colliding-wind
shock is much closer to the white dwarf, and/or the star produces many fewer ionizing photons.  Further observations
as suggested in Section \ref{sec:future} may reveal which process, or how much of each, is responsible
for the ionized gas.

%In the analysis above we have not made a geometrical or dynamic model of the radio emission regions in SySts.
%The numbers derived on Table {tab:results} are based on a simple uniform slab of ionized gas, it would be
%difficult to improve on the generic models of \citet{Seaquist_etal_1984} and \citet{Taylor_Seaquist_1984}
% without maps of the ionized region with resolution of a few tens of mas.  But the wide range of values for
%the luminosity, emission measure, size, and electron density resulting from the simple calculations above
%suggest that 

\acknowledgements {This research has made use of of NASA's Astrophysics Data System; the SIMBAD database and VizieR catalogue access tool, CDS, Strasbourg, France; and matplotlib for python \citep{matplotlib}.

J.L.S. was supported by NSF award AST-1616646.

This research was supported in part by the Australian Research Council grants DP0770157 and DP0559613.

The authors are grateful to the referee for thorough and helpful reviews of the manuscript.

The Australia Telescope Compact Array is part of the Australia Telescope National Facility which is funded by the Australian Government for operation as a National Facility managed by CSIRO.  

This work has made use of data from the European Space Agency (ESA) mission
{\it Gaia} (\url{https://www.cosmos.esa.int/gaia}), processed by the {\it Gaia}
Data Processing and Analysis Consortium (DPAC,
\url{https://www.cosmos.esa.int/web/gaia/dpac/consortium}). Funding for the DPAC
has been provided by national institutions, in particular the institutions
participating in the {\it Gaia} Multilateral Agreement.
}

\appendix

\section{Making the Source Model More Realistic},
\label{sec:appendix}

\subsection{Varying the Electron Temperature}
\begin{figure}[h]
\hspace{1in}\includegraphics[width=5in]{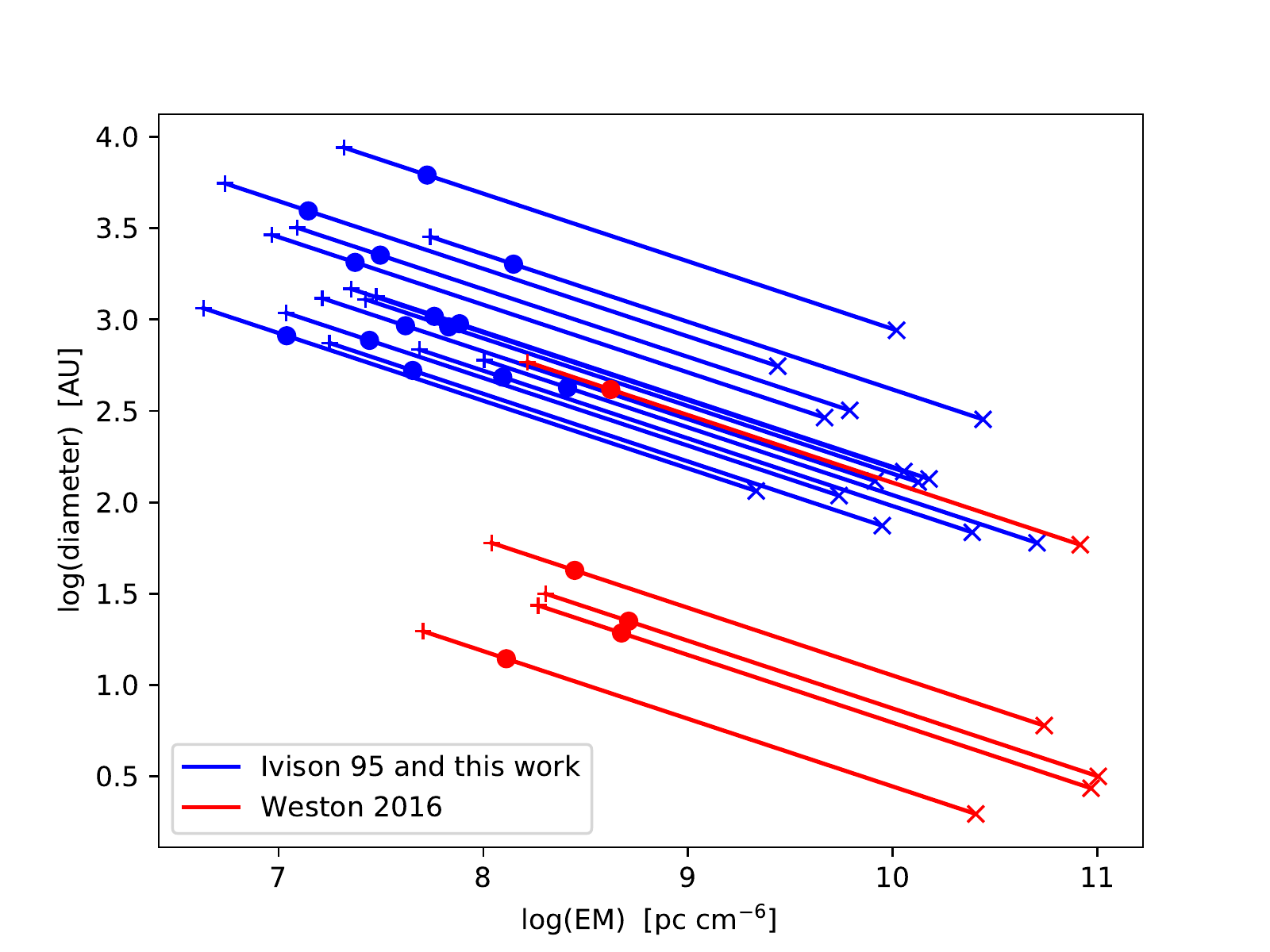}
\caption{Variation of the derived values of emission measure, $EM$, and
diameter with $T_e$, the assumed electron temperature.  The circles correspond
to the points on figure \ref{fig:alpha_4panel} and Table \ref{tab:results}, with $T_e = 10^4$ K.  The
lines show how those points would move for $5 \cdot 10^3 < T_e < 5\cdot 10^5$ K.
Decreasing $T_e$ by a factor of two reduces the derived value for
$EM$ and increases the diameter,
thus moving the point left and up to the $+$ position, while increasing $T_e$ by
a factor of 50 moves the point down and to the right as indicated by the $\times$
symbol.}
\label{fig:Temp_EM_size_2}
\end{figure}

One of the many interesting features of SySts' radio emission is the possibility
that it may come from regions of relatively high electron temperature, $T_e$,
as discussed in section \ref{sec:future}.
%The values of the brightness
%temperature computed in the lower right panel of figure \ref{fig:alpha_4panel}
%and the resulting values of $\Omega$, $d$, and $n_e$
%depend on the assumed value of $T_e = 10^4$ K.
Changing the assumed value of $T_e$ changes
both the emission measure, $EM$, derived from $C_1$ in equation \ref{eq:C1},
as well as the brightness temperature derived by equation \ref{eq:TB}
and the resulting values of $\Omega$, $d$, and $n_e$.
%In this analysis, the values of $C_1$ and the optical depth $\tau_{\nu}$
%are determined directly from $\alpha$, so changing the assumed value of
%$T_e$ changes $T_B$ proportionately, and $EM$ changes as $T_{e}^{1.35}$.
The derived values of density and size change with $T_e$ as shown on figure
\ref{fig:Temp_EM_size_2}.  For each of the sources on Table
\ref{tab:results} the circle on figure  \ref{fig:Temp_EM_size_2} shows the
derived values of $EM$ and the log of the diameter (columns 7 and 10), with
lines showing how these values would change depending on the assumed value of
$T_e$.  Decreasing $T_e$ to $5\cdot 10^3$ K would move the points up and to the left
on the figure to the locations marked by $+$ symbols, while increasing $T_e$
to $5\cdot 10^5$ K moves the points down and to the right as indicated by
the $\times$ symbol.
The implied density $n_e \propto \sqrt{EM/d}$ increases as the electron temperature
increases.

The sources on Table \ref{tab:results} include X-ray sources of types $\alpha,\ \beta, \ \delta$, and $\beta$/$\delta$ in the classification system of \citet{Luna_etal_2013}.
Type $\beta$, for which the X-rays originate in a region of collision between winds
from the white dwarf and the red giant star, are most likely to show a correspondence
between the ionized gas regions emitting the radio and X-rays.  These include RX Pup,
BI Cru, AG Peg, He 2-104, UV Aur, ZZ CMi, and NQ Gem.  Whether or not these show
higher electron temperatures in the radio is a question that will require combined
modelling of data taken in the radio and X-ray bands simultaneously.

\begin{figure}[h]
%\hspace{1in}\includegraphics[width=3.5in]{spherical_n/Components_RX_Pup.pdf}
%\hspace{1in}\includegraphics[width=3.5in]{spherical_n/Components_BI_Cru.pdf}
%\hspace{1in}\includegraphics[width=3.5in]{spherical_n/Component_AB_nsv19500.pdf}
\hspace{1.5in}\includegraphics[width=4in]{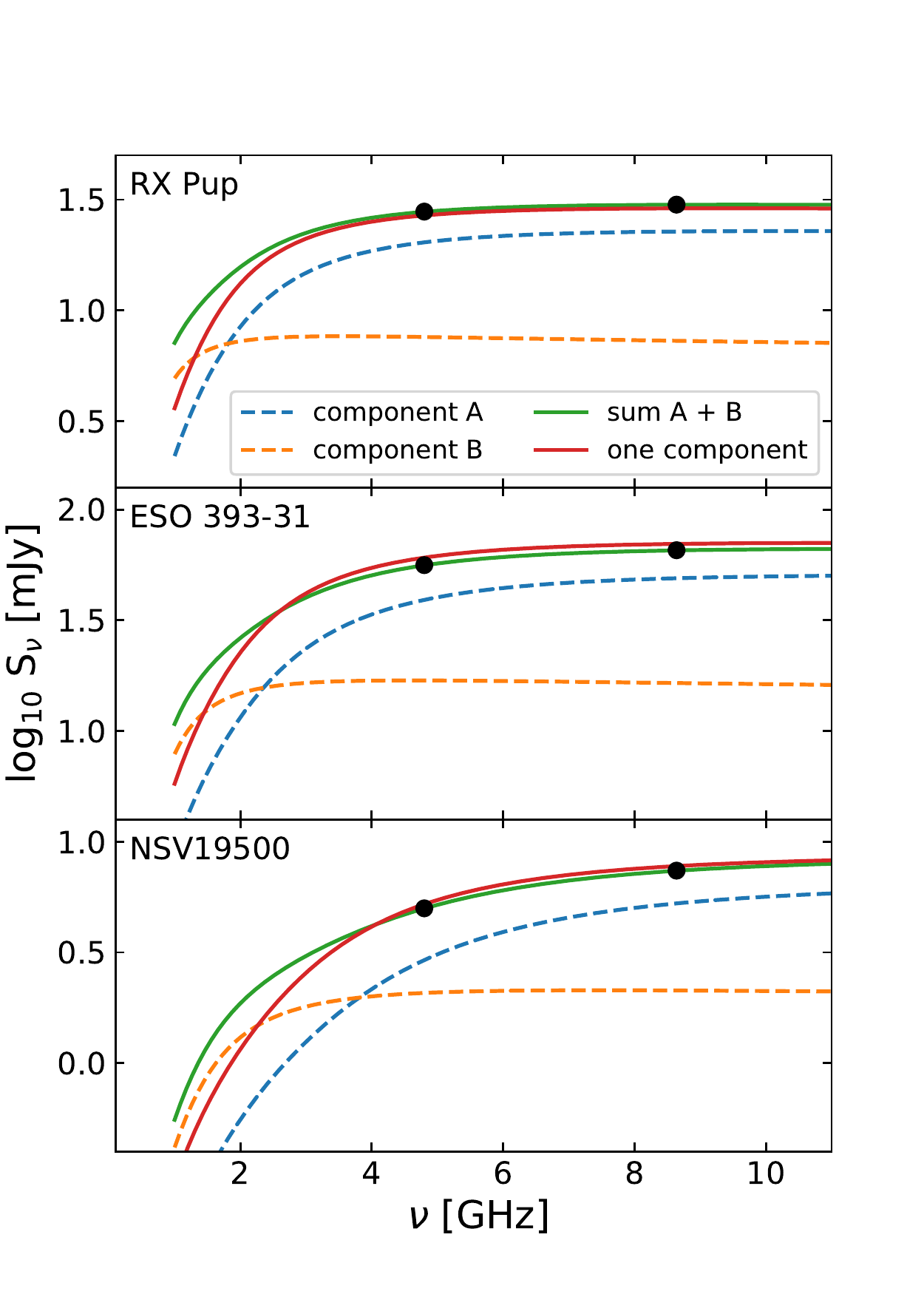}
\caption{Two component models of the density distribution.  The
%\textcolor{red}{
spectra in the top panel are for the low optical depth source, RX Pup ($\tau \simeq 0.11$).
The middle panel shows spectra for the intermediate optical depth source ESO 393-31 ($\tau \simeq 0.19$).
The bottom panel shows spectra for the higher optical depth source NSV 19500 = SS 73 38 ($\tau \simeq 0.44$).
The black dots show the measured flux densities at 4.8 and 8.64 GHz that constrain the model parameters.
The solid red curve is the best fit single component fit, as given on Table \ref{tab:results}, and as $EM_1$
on Table \ref{tab:2comp}.  The dashed lines show the contribution from the two components separately,
and the solid green line is their sum.}
%}
\label{fig:2comp}
\end{figure}

\subsection{Two Components with Different Emission Measures}

The results on table \ref{tab:results}, columns 6 - 11, assume a cylindrical geometry for the source,
so that the emission measure has the same value for all lines of sight through the
ionized gas.  This assumption allows derivation of representative values
for the physical parameters of the source, but it is far from realistic as a density distribution.
A somewhat more robust model would allow the emission measure to vary with position, e.g. in a
core-halo structure that might represent an ionized wind around a central high density region.
Since almost all SySts are unresolved by the ATCA, we cannot separate the contributions
of the different regions to the overall emission sprectrum.  But we can get a rough idea of the variation
in the physical parameters for a more complex density structure by making a two-component model
of the emission measure, $EM$, and the solid angle, $\Omega$.

If we assume that the emission from the source comes from two regions, one (component A) with emission measure ten times greater
than the other (component B), but with solid angle just one third as large, then we can make an estimate of how much the
results on Table \ref{tab:results} would have to change.  These ratios of $\frac{EM_A}{EM_B} = 10$ and $\frac{\Omega_A}{\Omega_B} = \frac{1}{3}$
mean that at low frequencies, where both components are optically thick, the emission comes mostly (75\%) from component B,
but at high frequencies where both components are optically thin, the emission comes mostly (77\%) from component A, since its
optical depth and hence brightness temperature will be ten times higher than $T_B$ in region B.
At the X-band frequencies of this study, most sources are in between these two limits, as illustrated on figure \ref{fig:2comp}.
Given the two measured flux densities at two frequencies (4.8 and 8.64 GHz), and the two assumed ratios for $EM$ and $\Omega$
between the two components, we can solve a system of four simultaneous equations to find both values of emission measure, $EM_A$ and $EM_B$, both
values of solid angle, $\Omega_A$ and $\Omega_B$, and from these the values of all the other parameters on Table \ref{tab:results} for
both components.

%\textcolor{red}{
Shown on figure \ref{fig:2comp} are three sources with quite different values of $\alpha$ between 4.8 and 8.64 GHz:  RX Pup that is very
optically thin ($\alpha \simeq 0.11$), ESO 393-31 that is moderately optically thin ($\alpha \simeq 0.19$), and the higher optical depth
source NSV 19500 = SS 73 38 ($\alpha \simeq 0.44$).%} 
Spectra for the two components (A and B) are shown on each panel, their sum
spectrum is shown, and for comparison the single component spectrum corresponding to the values from table \ref{tab:results}.
The corresponding results for components A, B, and the one component model (1) are summarized on table \ref{tab:2comp} for the
three sample stars.

\begin{table}[h]
\centering
%\textcolor{red}{
\begin{tabular}{|c|c|c|c|} \hline
%Star & RX Pup & BI Cru & NSV19500 \\
Star & RX Pup & ESO 393-31 & NSV 19500 \\
\hline
$\log_{10}{(EM_A)}$ & 7.6  & 7.9 & 8.3 \\
$\log_{10}{(EM_B)}$ & 6.6 & 6.9 & 7.3 \\
$\log_{10}{(EM_1)}$ & 7.5 & 7.7 & 8.1 \\
\hline
$\log_{10}{(\Omega_A)}$ arcsec$^2$ & -0.5 & -0.4 & -1.7 \\
$\log_{10}{(\Omega_B)}$ arcsec$^2$ & -0.04 & +0.08 & -1.2 \\
$\log_{10}{(\Omega_1)}$ arcsec$^2$ & -0.3 & -0.1 & -1.4 \\
\hline
$\tau_A$ (8.64 GHz) & 0.15 & 0.26 & 0.71 \\
$\tau_B$ (8.64 GHz) & 0.015 & 0.026 & 0.071 \\
$\tau_1$ (8.64 GHz) & 0.11 & 0.19 & 0.44 \\
\hline
diameter $d_A$ (AU$\cdot10^2$) & 17  & 46  & 3.4 \\
diameter $d_B$ (AU$\cdot10^2$) & 30  & 80  & 5.9 \\
diameter $d_1$ (AU$\cdot10^2$) & 22  & 62  & 4.8 \\
\hline
$\log_{10}{(n_{e,A})}$ & 4.9 & 4.8 & 5.5 \\
$\log_{10}{(n_{e,B})}$ & 4.1 & 4.1 & 4.9 \\
$\log_{10}{(n_{e,1})}$ & 4.7 & 4.6 & 5.4 \\
\hline
\end{tabular}
%}
\caption{Physical parameters of the two component emission model, compared with the one component results, for the
three representative sources illustrated on figure \ref{fig:2comp}. The units of $EM$ are pc cm$^{-6}$ and $n_e$ is in cm$^{-3}$.}
\label{tab:2comp}
\end{table}

%\textcolor{red}{
The two component results shown on table \ref{tab:2comp} are typically different by factors of three to six, except for the emission
measures, that are forced to have a ratio of ten, and the optical depths, that must also differ by a corresponding factor of ten.  In most cases  the one component results are
roughly the geometric mean of the results for components A and B, thus different
from each by about a factor of two.  
%An exception is the diameter of the source, where the one component results are larger than either of components A or B, but less than their sum.  
Although we can imagine that component B would be concentric with component A, as a halo
around the higher density core, that is not assumed here.  As long as they are both contained within the beam, their juxtaposition
does not matter.  The result of this analysis is to suggest that a factor of three variation in the sizes and electron densities on
table \ref{tab:results} might be expected due to source structure, but a factor of ten or more would be unlikely.%} 
Of course, other
source components, e.g. a low density wind,  that do not contribute significantly to the flux at either frequency, are not constrained
by these observations.

\vspace{.2in}

\subsection{ %\textcolor{red}{
Spatial Structure of the Electron Density}%}

\begin{figure}[h]
\hspace{1in}\includegraphics[width=5in]{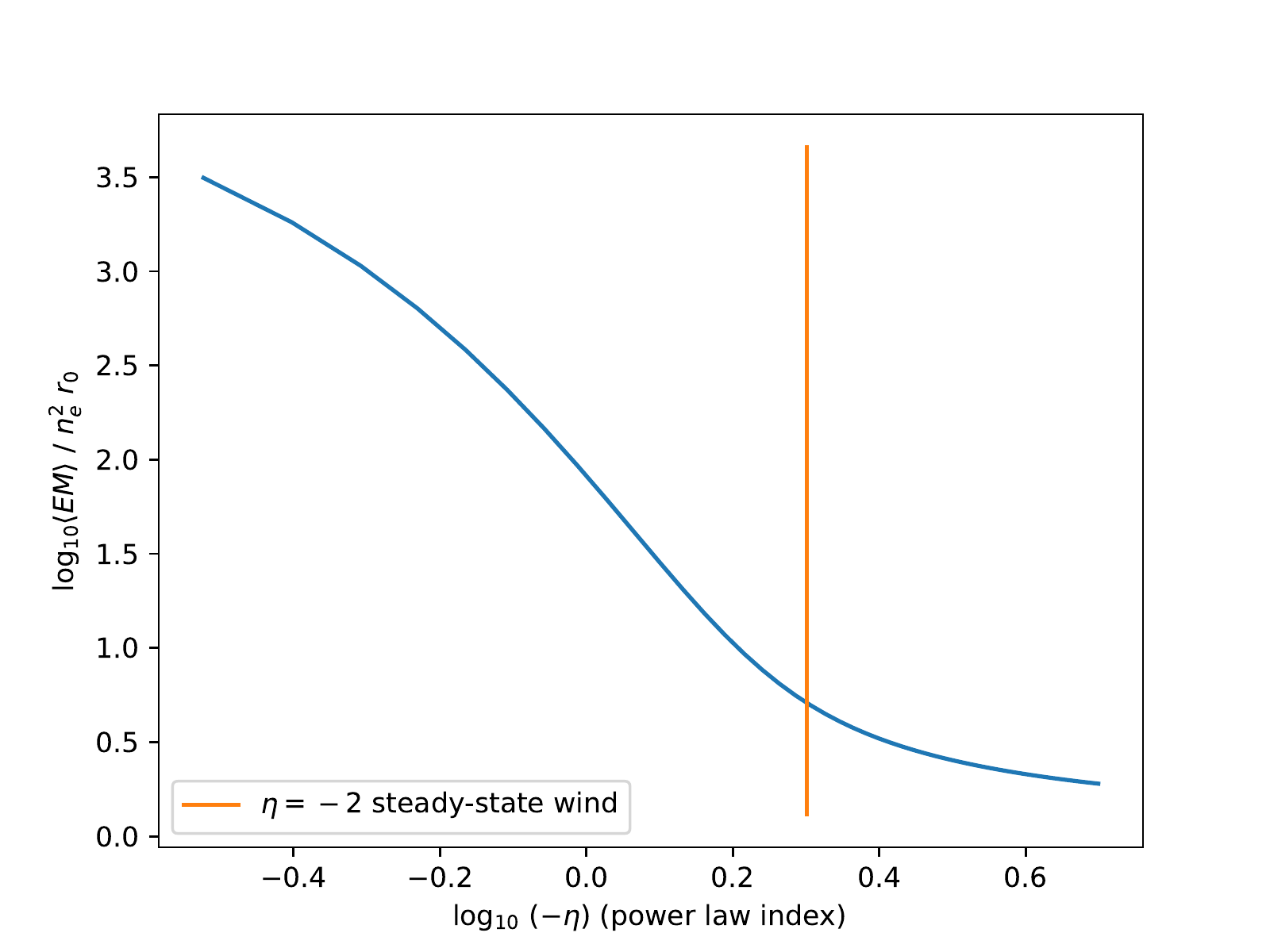}
\caption{Variation of the mean emission measure $\left<EM\right>$,
equation \ref{eq:wind2}, with the
power law index, $\eta$, of a steady-state wind,
equation \ref{eq:wind}.
For a cylinder $\left<EM\right> = n_e^2 \ d $,
for the steady-state wind $\left<EM\right> \simeq 5.1\ n_0^2\ r_0$.
}
\label{fig:wind}
\end{figure}

The values on the last five columns of table \ref{tab:results}
assume not only that $T_e = 10^4$ K but also that the
density distribution is a uniform cylinder, with length
along the line of sight, $s$, equal to its diameter, $d$,
so that the emission measure is simply $EM = n_e^2 d$
that is found from $C_1$ by equation \ref{eq:C1}.
For a sphere of uniform density the $EM$
%solid angle is also (pi/4) (d/D)**2, but the EM
changes with position, from a peak at
  the center, to zero at the edge, as
$EM(b) = n_e^2 \sqrt{ (d)^2 - (2b)^2}$ with $b$ the
impact parameter, or minimum distance from the center to the line of sight.  Averaged over the solid angle of the source, the mean $EM$ of a sphere is $\left<EM\right> = \frac{2}{3}\ d\ n_e^2$.
For an unresolved source with a given value of $C_1$ (derived from the measured value of $\alpha$ by equation \ref{eq:alpha_formula}), the sphere has lower $\left<EM\right>$ than the cylinder by a factor of 2/3.
So to explain the observed $\alpha$ the diameter must by
larger than the value on table \ref{tab:results}, column
10 by a factor of 1.5, or the density must by
larger than the value in column 11 by a factor of 1.22.

A more realistic distribution of $n_e$ in the SySt environment
is a steady state wind, with constant velocity and
mass-loss rate, so that the density is
\begin{equation}\label{eq:wind}
\frac{n_e}{n_0}\ = \ \left(\frac{r}{r_0}\right)^{\eta}
\end{equation}
with power law index $\eta = -2$.  The
wind starts at radius $r_0$, within which the density is a constant, $n_0$.
The mass and solid angle diverge for this wind, but the flux
density and the solid angle average of the
emission measure are finite.  Integrating $n_e^2$ along
lines of sight, $\int ds$, at a range of values for $b$ gives
\begin{equation}\label{eq:wind2}
\left<EM\right>\ = \ \frac{\int \ 2 \pi b \left[ \int\ n_e^2(b,s) ds \ \right]\ db}{\pi r_0^2} \ \simeq \ 2.6 \ n_0^2 \  d
\end{equation}  where $d = 2 r_0$ and $n_e(b,s) = n_e(r)$ given by equation \ref{eq:wind}
with $r = \sqrt{b^2 + s^2}$, and $s$ is the line of sight distance from the point of
closest approach to the center ($s=0$).
For the $\eta = -2$ wind, $\left<EM\right>$ is increased relative to the hard-edge sphere by a factor of about 3.8, and relative to the cylinder with diameter $d$ the wind
has $\left<EM\right>$ higher by a factor of $\sim$2.6.
This would lead to values for $d$ lower by a factor of 2.6
relative to the numbers on table \ref{tab:results}, column
10, or values for $n_e$ lower by $\sqrt{2.6}\simeq 1.6$,
i.e. a decrease of 0.2 in $\log_{10}{(n_e)}$
on column 11 %\textcolor{red}{
or a combination of a smaller change in both quantities.%}

For different power-law indices $\eta$ in equation \ref{eq:wind}
the $\left<EM\right>$ varies with
$\eta$ as shown in figure \ref{fig:wind}.  The special case of a steady-state, constant velocity wind
($\eta = -2$) is indicated by the vertical line.  Other values of $\eta$
require either acceleration/deceleration of the wind,
e.g. due to a shock, which is typical of SySts with X-ray spectra of type $\beta$ \citep{Luna_etal_2013},
or variations in the mass loss rate with time, as is seen in the radio emission shortly after
a nova outburst \citep[e.g.][]{Weston_etal_2016}.
More comprehensive and informative models for the radio emission regions of SySts
take into account the rich detail of the optical, IR, and X-ray spectra
as well as long and short term changes in their luminosity in all
bands \citep[e.g.][]{Contini_etal_2009}.

The density models discussed in this section allow simple integration of $n_e^2$
to compute the mean emission measure, $\left<EM\right>$, that is derived from
$\alpha$ by equations \ref{eq:C1} and \ref{eq:alpha_formula} for an unresolved source.
If the sources were well resolved by the telescope, then the computation would be
much more complicated, because the source might be optically thin in some parts
($\alpha \simeq -0.1$), optically thick in others ($\alpha \simeq +2$).  For those
limiting values, the error in a measurement of $\alpha$ leads to a very large error
in $\tau$, and the derived value of the $EM$ is very imprecise, i.e. the
horizontal parts of the curves in the upper right panel of figure
\ref{fig:alpha_4panel}.  The point of the
calculations in this section is that, for unresolved sources
where $\alpha$ and hence $\tau$
are fairly well determined, the resulting $\left<EM\right>$ and hence the size and density of the
emission region are close to the values given by the uniform cylinder model (Table
\ref{tab:results}).

%\section{Ultra-violet Fluxes of Selected Sources}\label{sec:uvot}
%
%For the sources observed as part of the June-July 2007 campaign (table \ref{tab:positions}), we have collected ultraviolet flux densities using the UVOT instrument on
%SWIFT ref \citep{???}, given on table \ref{tab:uvot}.  The UVOT observations 
%are not contemporary with the radio data, in many cases they come from several
%years after. The MJD of the UVOT observations are
%given in column 4 of table \ref{tab:uvot}, while the radio data are from MJD 55280
%-- 55286.  We do not find a correlation between the radio flux densities and the
%UVM2 flux densities (figure \ref{fig:uvot}).  Since
%the uv flux is very sensitive to dust obscuration,
%while the radio is not, the lack of correlation cannot be taken as evidence
%against UV ionization from the white dwarf star as the source of energy
%for the radio emission.
%
%\input{Swift_uvot_2007_symb_tab.tex}
%
%\begin{figure}[h]
%\hspace{2in} \includegraphics[width=4in]{Radio_vs_UVM2.pdf}
%\caption{UV fluxes in UVOT band UVM2 ($\nu = 1.345 \cdot 10^{15}$ Hz)
%vs. radio flux densities at 4.8 GHz.  Note that the radio emission
%from ESO 393-31 is dominated by the planetary nebula.  The data were 
%taken at different epochs.}
%\label{fig:uvot}
%\end{figure}

\end{document}